# Pulse-Width Modulation Neuron Implemented by Single Positive-Feedback Device

Sung Yun Woo, *Member, IEEE*, Dongseok Kwon, Byung-Gook Park, *Fellow, IEEE*, Jong-Ho Lee, *Fellow, IEEE* and Jong-Ho Bae, *Member, IEEE*

*Abstract*—Positive-feedback (PF) device and its operation scheme to implement pulse width modulation (PWM) function was proposed and demonstrated, and the device operation mechanism for implementing PWM function was analyzed. By adjusting the amount of the charge stored in the $n^-$ floating body ($Q_n$), the potential of the floating body linearly changes with time. When $Q_n$ reaches to a threshold value ($Q_{th}$), the PF device turns on abruptly. From the linear time-varying property of $Q_n$ and the gate bias dependency of $Q_{th}$, fully functionable PWM neuron properties including voltage to pulse width conversion and hard-sigmoid activation function were successfully obtained from a single PF device. A PWM neuron can be implemented by using a single PF device, thus it is beneficial to extremely reduce the area of a PWM neuron circuit than the previously reported one.

## I. INTRODUCTION

As a result of recent interest in deep neural networks (DNNs), DNNs show similar level of inference performance to human for specific tasks, such as recognizing hand-written digits and classifying certain images [1]-[3]. However, the more complex the work required for DNN, the more complex the structure of the DNN and the more computations required. Hardware-based neural networks (HNNs) efficiently perform large-scale parallel multiply-accumulation (MAC) operations based on Ohm's and Kirchhoff's laws, and exponentially save power consumption compared to conventional computing systems [4]-[7]. There has been a lot of researches to utilize memory devices, such as resistive change memory (RRAM), phase change memory (PCRAM), and charge trap devices like FLASH [8]-[13], to represent synaptic weights and performing MAC operation. For MAC operation using the memory array, input data which is a set of raw data or the results of a MAC operation should be coded into a different form of signal by using a neuron circuit [14]. There are three mainstreams of the ways to input coding: pulse amplitude modulation (PAM), pulse width modulation (PWM), and pulse rate modulation (rate coding) [15]-[17]. PAM is simplest way for data coding, but it is inadequate for HNN because of the nonlinear output characteristics of most synaptic devices. Rate coding and PWM use voltage pulse with constant amplitude, thus those are beneficial in terms of accurate MAC operation [18].

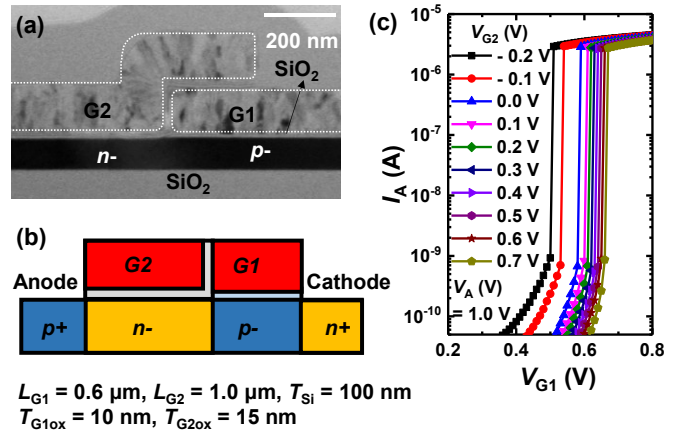

Fig. 1 (a) A TEM image and (b) a schematic cross-sectional view of the PF device. (c) The measured $I_A$-$V_{G1}$ characteristics of the fabricated PF device at different $V_{G2}$, and $V_A$ of 1 V.

Previously, we simulated an HNN using SPICE with a pulse-width modulation (PWM) neuron circuit consists of 13 transistors and one capacitor (13T-1C) [18]. It is noteworthy that the neuron circuit should be able to be integrated in a small area, considering that an HNN requires a very large number of neurons.

In this paper, a positive-feedback (PF) device and the novel operation scheme for implementing a PWM neuron is proposed. The PWM function was successfully demonstrated using a single PF device. Considering the size and the number of transistors for the previously reported PWM neuron circuit, the size of the PWM neuron can be extremely reduced with a single PF device. In addition, the voltage amplitude to pulse width conversion can be performed without capacitor in the PF device, thus the PWM neuron size would be reduced more.

## II. DEVICE STRUCTURE AND POSITIVE FEEDBACK OPERATION

PF devices have been used to implement neuronal functions [19]-[22], thanks to their unique behavior. Fig. 1(a) and (b) show the cross-sectional schematic and SEM images of the PF device used in this work. The detailed fabrication step is described elsewhere [19]. The device has $p^+/n^-/p^-/n^+$ doped

This work was supported in part by the (ERC) and the National Research Foundation of Korea (NRF- 2016M3A7B4909604). The review of this letter was arranged by Editor XXX. (Corresponding authors: Jong-Ho Lee; Jong-Ho Bae.) S. Y. Woo, D. Kwon, B.-G. Park and J.-H. Lee are with the Department of Electrical and Computer Engineering and Inter-University Semiconductor Research Center (ISRC), Seoul National University, Seoul 08826, Korea (e-mail: jhl@snu.ac.kr).
  J.-H. Bae is with the School of Electrical Engineering, Kookmin University, Seoul 02707, Korea (e-mail: jbae@kookmin.ac.kr)..

channel and two gates (G1 and G2) located next to each other on $p^-$- and $n^-$-doped region. The doping concentrations of $n^-$/$p^-$ channel are $1\times10^{18}$ and $3\times10^{17}$ cm$^{-3}$, respectively. The lengths of G1 and G2 are 0.6 μm and 1.0 μm, respectively, and the thickness of $p^+$/$n^-$/$p^-$/$n^+$ doped channel is 100 nm. Gate insulator for G1 is SiO$_2$ with thickness of 10 nm and for G2 has EOT of 15 nm. This PF device is basically two connected floating body MOSFETs: n- and p-MOSFET having G1 and G2 as a gate, respectively. These MOSFETs are connected in series and their drains are respectively connected to the bodies of each other. When both the MOSFETs are turned-off, applying a positive anode bias ($V_A > 0$) with grounded cathode ($V_C = 0$) cannot turn on the device due to the reverse-biased p-n diode formed by $n^-$/$p^-$ junction. However, when a positive G1 bias ($V_{G1}$) is applied, which is a gate bias of n-MOSFET, electrons are injected from the cathode to the $n^-$ region and stored, which is the drain of the n-MOSFET and the body of the p-MOSFET. In that case, the threshold voltage of p-MOSFET($V_{Tp}$) increases due to the floating body effect. As $V_{Tp}$ increases, the amount of the hole injection from anode to p-region exponentially increases and it also reduces the threshold voltage of n-MOSFET ($V_{Tn}$), therefore, more holes are injected from the anode to the $p^-$ region. In summary, the drain current of n-MOSFET ($I_{Dn}$) injects electron and the stored electron in $n^-$ floating body ($Q_n$) change $V_{Tp}$ to increase $I_{Dp}$, thus p-MOSFET turns on and a positive feedback loop is formed. When $|Q_n|$ is larger than a threshold value ($Q_{th}$), the device is turned on and operates as a diode. Fig. 1(c) shows the anode current ($I_A$) versus $V_{G1}$ as a function of G2 bias ($V_{G2}$). Due to the positive-feedback operation, $I_A$ changes abruptly as $V_{G1}$ changes. It is worth noting that the turn-on $V_{G1}$ ($V_{G1,on}$) increases as $V_{G2}$ increases. As $V_{G2}$ increases, $V_{Tp}$ should be larger to turn-on the PF device, which means larger $Q_{th}$ is required.

## III. PULSE-WIDTH-MODULATION (PWM) OPERATION

The previously proposed PWM circuit consists of three functional circuit elements, a sawtooth wave generator, a differential amplifier and a level shifter (see Fig. 2(a)). The two inputs of

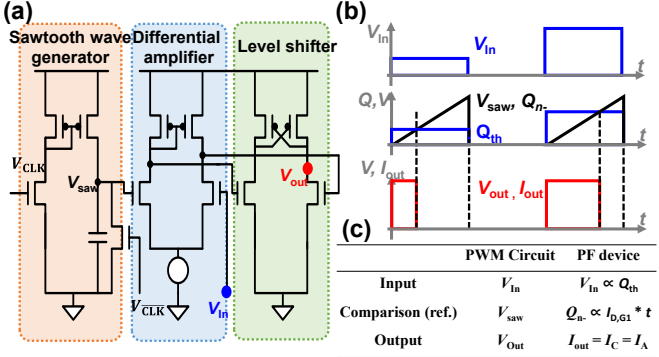

Fig. 2 (a) A schematic diagram of a PWM circuit [F1] (reproduced from [F1] with permission). (b) Schematic diagram of elements for implementing PWM function. (c) The functional element of a PF device for implementing PWM function.

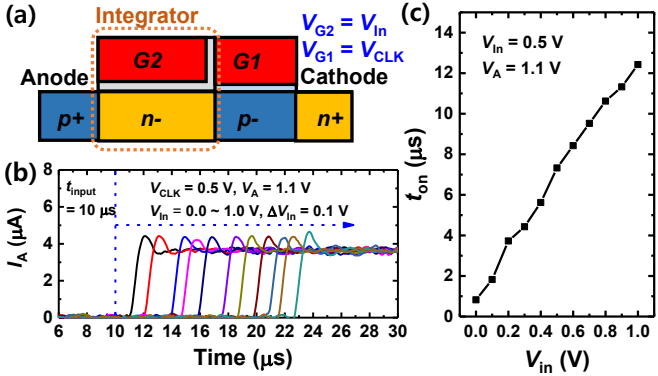

Fig. 3 (a) A bias scheme for measuring the PF device. (b) The measured transient responses ($t$–$I_A$) of the PF device under different $V_{in}$ and (c) the relationship between $V_{in}$ and $t_{on}$.

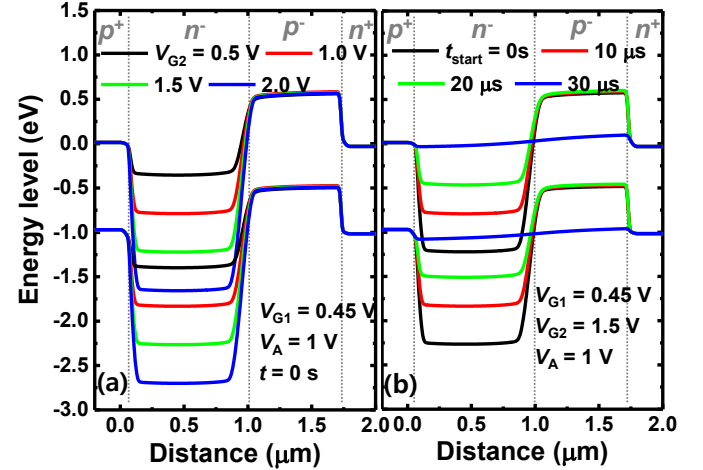

Fig. 4 (a) Energy band diagrams of a PF device along the silicon channel at different $V_{G2}$s (=$V_{in}$s). (b) Energy band diagrams of a PF device along the silicon channel over time, at $V_{G1} = 0.45$ V, $V_{G2} = 1.5$ V and $V_A = 1$ V.

the differential amplifier are the input data coded in a constant voltage amplitude ($V_{in}$) and a triangular voltage pulse train ($V_{saw}(t)$) generated by the sawtooth wave generator. In the sawtooth wave generator, a signal of $V_{saw}(t) = I/C_{saw}(t - t_0)$ is generated when clock (CLK) is 1, where $t_0$ is the time when CLK becomes 1 and the voltage increasing rate is the ratio between the pull-up current of the p-MOSFET($I$) and the capacitance ($C_{saw}$). As shown in Fig. 2(b), if $V_{in} > V_{saw}(t)$, the result of the differential amplifier ($V_{Diff}$) is 1 and if not, $V_{Diff}$ is 0. Considering $V_{saw}(t)$ is linear function of the time, the time for $V_{in} > V_{saw}(t)$ and the pulse width ($t_p$) of $V_{diff}$ are linearly proportional to $V_{in}$. Note that $V_{saw} = V_{DD}$ and $V_{saw}(t) > V_{in}$ when CLK is 0, thus maximum $t_p$ ($t_{max}$) is $1/f_{CLK}$ where $f_{CLK}$ is the clock frequency. Therefore, $V_{in}$ is linearly converted to a $V_{Diff}$ pulse with $t_p$ in the range of [0, $t_{max}$]. $V_{Diff}$ pulse is shifted to a desired voltage level in the level shifter as an output voltage pulse ($V_{out}$), and is used as a signal for the next layer. To summarize, the basic principle to implement PWM function is comparing a input signal ($V_{in}$) and a signal ($V_{saw}(t)$) that varies linearly with time during a clock cycle. Those can be realized by a single PF device as described in Fig. 2(b) and (c), which will be discussed in the following.

## IV. Implementation of PWM Function Using a Single PF Device

The PWM function was implemented from a fabricated PF device by applying $V_{G1}$ and $V_{G2}$ as a CLK bias ($V_{CLK}$) and $V_{in}$, respectively, as shown in Fig. 3. As $V_{G2} = V_{in}$ increases, the time to turn-on the PF device ($t_{on}$) increases (Fig. 3(b)), and $V_{in}$ and $t_{on}$ has a linear relationship (Fig. 3(c)). This result implies that the fabricated PF device can operate as a linear $V_{in}$–$t_p$ converter in a certain range.

In order to analyze the origin of the linear relationship between $V_{in}$ and $t_{on}$ which is a key mechanism of implementing PWM function, a simulation study was conducted with a TCAD simulator (Sentaurus) of Synopsys (Figs. 4 and 5). As $V_{G2} = V_{in}$ increase, the energy band at $n^-$ region linearly moves down and $V_{Tp}$ linearly decreases (Fig. 4(a)), because the $n^-$ region is floated and is deep-depleted. The key is that the degree of deep-depletion is linearly proportional to the $V_{G2} = V_{in}$, and $Q_{th}$ linearly increases. Considering $V_{G1} = V_{CLK} = 0.45$ V and assuming the source of $Q_n$ is $I_{Dn}$, $|Q_n|$ increases linearly proportional to the time and $|Q_n| = I_{Dn}(t - t_0)$. Thus, the energy band at $n^-$ region moves upward as time goes by, as shown in Fig. 4(b). In terms of implementing PWM function, $Q_{th}(V_{in})$ at $t = t_0$ is the constant input signal and $|Q_n(t)|$ is the linear time-varying signal (see Fig. 2(b) and (c)). These two signals are compared whether $Q_{th}(V_{in}) > |Q_n(t)|$ or not. When $Q_{th}(V_{in}) > |Q_n(t)|$, p-MOSFET does not turn-on. When $Q_{th}(V_{in}) < |Q_n(t)|$, p-MOSFET turns on and the PF device turns on due to the positive feedback operation, as explained in Section II.

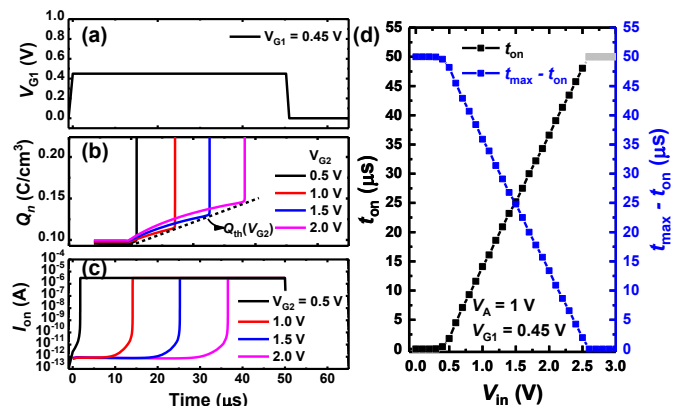

Fig. 5 (a) $V_{CLK}$, (b) $Q_n$ and (c) $I_A$ over time at various $V_{G2}$ (=$V_{in}$). (d) $V_{in} - t_{on}$ (left y-axis) and $V_{in} - (t_{max} - t_{on})$ (right y-axis) plots which indicates activation function of the proposed PWM function implemented by a PF device.

Fig. 5(b) shows $|Q_n(t)|$ where $V_{G2} = V_{in}$ varies from 0.5 V to 2.0 V, when $V_{G1} = V_{CLK} = 0.45$ V (see Fig. 5(a)). As $V_{G2} = V_{in}$ increases, $Q_{th}$ and $t_{on}$ linearly increases, as shown in Fig. 5(b) and (c). Therefore, a linear relationship between $V_{G2} = V_{in}$ and $t_{on}$ is obtained. PF device is off ($I_A \sim 1$ pA) when $V_{CLK} = 0$, thus $t_{max}$ is $1/f_{CLK}$. This result indicates $V_{in}$ is successfully converted to $t_{on}$ with linear relationship in the range of [0, $t_{max}$], and the pulse width is $t_{max} - t_{on}$. The relationship between $V_{in}$ and $t_{on}$ is the hard-sigmoid activation function between 0 and $t_{max}$, and it can be modulated by using additional circuitry, such as a simple inverter. Therefore, the proposed PF device with this operation scheme is a PWM neuron that effectively implements both a $V_{in}$ to $t_p$ converter and a hard-sigmoid activation function.

## V. Conclusion

PWM function is implemented by using a single PF device with two gates. $V_{G1}$ is $V_{CLK}$ which is linearly increase $|Q_n|$ when CLK = 1, and $V_{Tp}$ linearly increases as time goes by. By applying input signal to $V_{G2}$, $Q_{th}$ is linearly changed and comparison between $|Q_n|$ and $Q_{th}$ is performed, and the result is amplified by the positive feedback operation. The proposed positive feedback device effectively integrates the two basic function elements of PWM circuits: generating linear time-varying variable ($|Q_n|$) and comparing it with input signal ($Q_{th}$). Therefore, the PWM circuit including a capacitor can be replaced by a single positive feedback, which is beneficial to integrate neuron circuits for neural network area.